\documentclass[10pt, preprint,authoryear]{elsarticle}

\usepackage{fullpage}
\usepackage{amsmath, amssymb, amsthm, bm}
\usepackage{dsfont}
\usepackage{float}
\newtheorem{proposition}{Proposition}
\newtheorem{lemma}{Lemma}

\newtheorem*{restatelemma}{\restatelemmaname}
\newcommand{\restatelemmaname}{}
\newenvironment{restatelemmawithname}[1]
  {\renewcommand{\restatelemmaname}{Lemma #1}%
   \begin{restatelemma}}
  {\end{restatelemma}}

\usepackage[ruled,vlined]{algorithm2e}
\usepackage{graphicx}
\usepackage{changepage}
\usepackage[
    colorlinks=true,   % turn off the funky link boxes
    linkcolor=blue,    % internal links (e.g., \ref)  
    citecolor=blue,    % citations
    urlcolor=blue      % URLs
]{hyperref}
\newcommand{\iidsim}{\overset{\textnormal{i.i.d.}}{\sim}}
\newcommand{\indsim}{\overset{\textnormal{ind}}{\sim}}
\newcommand{\independent}{\perp\!\!\!\perp}
\newcommand{\indicator}{\mathds 1}
\newcommand{\naive}{{\textnormal{naive}}}
\newcommand{\wtd}{{\textnormal{wtd}}}
\newcommand{\wtdp}{{\textnormal{wtd+}}}
\newcommand{\E}{\mathbb{E}}
\newcommand{\Var}{\operatorname{Var}}
\newcommand{\Cov}{\operatorname{Cov}}
\newcommand{\plim}{\mathop{\mathrm{plim}}}

\usepackage[left=0.75in, right=0.75in, top=0.75in, bottom=1in]{geometry} % TODO: delete 

\begin{document}

\title{\vspace*{0.25in} Location Tests with Noisy Proxies for Latent Variables}

\author[upenn]{Louis Deutsch}

\cortext[cor1]{Corresponding author.}

\author[upenn]{Eugene Katsevich\corref{cor1}}
\ead{ekatsevi@wharton.upenn.edu}

\address[upenn]{Department of Statistics and Data Science, University of Pennsylvania, Philadelphia, PA 19104, USA}
\date{April 14, 2025}

\begin{abstract}
\noindent
We investigate inference in a latent binary variable model where a noisy proxy of the latent variable is available, motivated by the variable perturbation effectiveness problem in single-cell CRISPR screens. The baseline approach is to ignore the perturbation effectiveness problem, while a recent proposal employs a weighted average based on the proxies. Our main goals are to determine how accurate the proxies must be in order for a weighted test to gain power over the unweighted baseline, and to develop tests that are powerful regardless of the accuracy of the proxies. To address the first goal, we compute the Pitman relative efficiency of the weighted test relative to the unweighted test, yielding an interpretable quantification of proxy quality that drives the power of the weighted test. To address the second goal, we propose two strategies. First, we propose a maximum-likelihood based approach that adapts the proxies to the data. Second, we propose an estimator of the Pitman efficiency if a ``positive control outcome variable'' is available (as is often the case in single-cell CRISPR screens), which facilitates an adaptive choice of whether to use the proxies at all. Our numerical simulations support the Pitman efficiency as the key quantity for determining whether the weighted test gains power over the baseline, and demonstrate that the two proposed adaptive tests can improve on both existing approaches across a range of proxy qualities.
\end{abstract}

\maketitle

\section{Introduction}

\subsection{Problem setup}

Latent variable models are commonly encountered and well understood. However, there is a phenomenon that has received less attention in this context: in certain applications, noisy proxies are available for the latent variables. For example, in single-cell CRISPR screens \citep{Dixit2016}, whether a genetic perturbation worked in a given cell is unobserved, but a proxy for this binary latent variable can be constructed by analyzing gene expression patterns in that cell \citep{Jiang2024}. Our goal is to study the implications of the availability of such noisy proxies for hypothesis testing in latent variable models. 

We consider a prototypical problem setting that captures the challenges and opportunities presented by noisy proxies for latent variables, a normal location test with binary latent variables:
\begin{equation}
Z_i \iidsim \textnormal{Ber}(\varphi); \quad Y_i \mid Z_i \indsim \mathcal N(\mu Z_i, 1); \quad \gamma_i \mid Z_i \indsim F_{Z_i}; \quad Y_i\independent \gamma_i \mid Z_i, \quad \text{for} \quad i = 1, \ldots, n.
\label{eq:outcome-model}
\end{equation}
Here, the binary latent variables $Z_i$ (loosely representing whether a perturbation worked in cell $i$) dictate whether the outcome variable $Y_i$ (loosely representing gene expression in cell $i$) is shifted by $\mu$ from its baseline value of 0, and $\gamma_i \in [0,1]$ is a continuous noisy proxy for $Z_i$. We avoid placing assumptions on the distribution $\gamma_i \mid Z_i$, and instead represent in non-parametrically by $F_{Z_i}$, as this distribution would be hard to learn in practice. Finally, the assumption $Y_i\independent \gamma_i \mid Z_i$ essentially states that the noisy proxies are obtained from a source independent of the outcome variables. In single-cell CRISPR screens, this may be obtained by holding out the gene of interest from the learning of the proxies $\gamma_i$, as proposed by \citet{Jiang2024}. Within the model~\eqref{eq:outcome-model}, the goal is to test the null hypothesis $H_0 : \mu = 0$ versus the alternative $H_1 : \mu > 0$ at level $\alpha$.

The simplest approach to testing $\mu = 0$ in the model~\eqref{eq:outcome-model} is to ignore the effect of the latent variable $Z_i$, employing a normal one-sample test. We term this the \textit{naive} test:
\begin{equation}
\hat{\mu}_\naive = \frac{1}{n} \sum_{i=1}^n Y_i; \quad \phi_\naive(\bm Y) = \indicator \left(\sqrt n \hat{\mu}_\naive > z_{1-\alpha}\right).
\label{eq:mu-hat-0}
\end{equation}
The naive test corresponds to the most common approach employed in the single-cell CRISPR screen analysis literature, e.g. as in \citet{Barry2023a} and \citet{Zhou2023a}. Recently, \citet{Jiang2024} proposed a method to learn proxies $\gamma_i$ and adjusted the naive test by using these proxies as weights. In the context of our simplified model~\eqref{eq:outcome-model}, this leads to the \textit{weighted} test:
\begin{equation}
\hat{\mu}_\wtd = \frac{\sum_{i=1}^n \gamma_i Y_i}{\sum_{i=1}^n \gamma_i}; \quad \phi_\wtd(\bm Y, \bm \gamma) = \indicator \left(\frac{\hat \mu_\wtd}{\text{s.e.}(\hat \mu_\wtd)} > z_{1-\alpha}\right), \quad \text{where} \quad \text{s.e.}(\hat \mu_\wtd) = \frac{\sqrt{\sum_{i = 1}^n \gamma_i^2}}{\sum_{i = 1}^n \gamma_i}.
\label{eq:mu-hat-1}
\end{equation}
It is easy to verify that both tests control Type-I error in finite samples. In particular, the validity of $\hat \mu_\wtd$ does not depend on the quality of the estimates $\gamma_i$, and is justified by conditioning on these quantities. This leaves power as the key consideration in the choice of test. The main questions we seek to address are as follows: (1) How accurately must the $\gamma_i$ approximate the $Z_i$ so that their incorporation into the weighted test results in a power increase over the naive test? (2) How can we conduct a powerful test without knowing the accuracy of the $\gamma_i$ for the $Z_i$?

\subsection{Related work}

If we view $\gamma_i$ as being a noisily measured version of $Z_i$, then our problem setup can be cast as a measurement error problem, which has been studied extensively \citep{Carroll2006a}. However, we depart from this literature by avoiding any modeling assumptions on the measurement error distribution $\gamma_i \mid Z_i$, which is possible due to our focus on testing rather than estimation. Our problem setting is also adjacent to that of testing for a treatment effect in a randomized experiment with noncompliance that is imperfectly observed \citep{Angrist1996, Frangakis2002, Boatman2018}, or in an observational study with an imperfectly observed exposure \citep{Kuroki2014, Cui2024, Zhou2024}. However, both of these strands of work focus on identification and estimation in the face of confounding due to noncompliance or measurement error. In our case, confounding is not a concern, and our focus is on testing and on resolving the power loss due to the ``noncompliance.''

\section{Comparing the power of the naive and weighted tests}
\label{sec:asymptotics}

In this section, we derive the asymptotic power of the naive~\eqref{eq:mu-hat-0} and weighted~\eqref{eq:mu-hat-1} tests against a sequence of local alternatives $\mu_n = h/\sqrt{n}$ in the model~\eqref{eq:outcome-model}, as well as the Pitman relative efficiency between the two tests. Although our focus is testing, rather than estimation, we begin our asymptotic analysis by determining the limiting distributions of the two estimators under local alternatives. This will provide insight into the behavior of the estimators underlying the tests and reveal the key properties of the data-generating process that govern their asymptotics. Before stating this result, we introduce a set of regularity assumptions that we will use throughout, all of which are mild:
\begin{equation}
\mu \in \Theta \equiv [-K, K] \text{ for some } K > 0; \quad 
\text{Var}[\gamma] > 0; \quad 
-\E[\log(1-\gamma)] < \infty.
\label{eq:assumptions}
\end{equation}
Now, we are prepared to state our first result. All proofs are deferred to \ref{sec:proofs}.
\begin{proposition}
\label{eq:naive-and-wtd-point-ests}
Consider a sequence of local alternatives $\mu_n = h/\sqrt{n}$ for some $h \geq 0$. Under the assumptions~\eqref{eq:assumptions}, we have
\begin{align}
\sqrt{n}(\hat \mu_\naive - \mu_n \varphi) &\xrightarrow{d} N(0, 1), \\
\sqrt{n}\left(\hat \mu_\wtd - \mu_n \varphi \frac{\mathbb E[\gamma \mid Z = 1]}{\mathbb E[\gamma]}\right) &\xrightarrow{d} N\left(0, \frac{\mathbb E[\gamma^2]}{\mathbb E[\gamma]^2}\right). 
\end{align}
\end{proposition}

The estimator $\hat\mu_\naive$ exhibits downward bias by a factor of $\varphi$ due to ignoring the effects of the latent variables $Z_i$, which echoes a classical phenomenon in measurement error models \citep{fuller2009measurement}. By contrast, $\hat \mu_\wtd$ compensates for this downward bias by a factor of $\mathbb E[\gamma \mid Z = 1]/\mathbb E[\gamma]$. This factor will be large to the extent that $\gamma$ is a good proxy for $Z$, because then $\mathbb E[\gamma\mid Z=1]$ will be large relative to the typical size of $\gamma$. However, the weighted estimator pays a price in terms of an increased variance over that of $\hat \mu_\naive$, since $\mathbb E[\gamma^2]/\mathbb E[\gamma]^2 \geq 1$.

Having quantified the asymptotic behavior of the estimators $\hat \mu_\naive$ and $\hat \mu_\wtd$, we can compute the local asymptotic power of the tests based on these estimators.

\begin{proposition}
\label{eq:naive-and-wtd-powers}
Under the assumptions~\eqref{eq:assumptions}, the asymptotic powers of the tests $\phi_\naive$ and $\phi_\wtd$ against the local alternatives $\mu_n = h/\sqrt{n}$ are given by 
\begin{align}
\lim_{n \rightarrow \infty} \mathbb E[\phi_\naive] &= 1 - \Phi\left(z_{1 - \alpha} - h\varphi\right); \\
\lim_{n \rightarrow \infty} \mathbb E[\phi_\wtd] &= 1 - \Phi\left(z_{1 - \alpha} - h\varphi \frac{\mathbb E[\gamma \mid Z = 1]}{\sqrt{\mathbb E[\gamma^2]}}\right).
\end{align}
Therefore, the Pitman efficiency of $\phi_\wtd$ relative to $\phi_\naive$ is given by
\begin{equation}
\psi^2 \equiv \frac{\mathbb E[\gamma \mid Z = 1]^2}{\mathbb E[\gamma^2]} = \frac{\left(\mathbb E[\gamma \mid Z = 1] / \mathbb E[\gamma]\right)^2}{1 + \textnormal{c.v.}^2[\gamma]},
\label{eq:pitman-efficiency}
\end{equation}
where $\textnormal{c.v.}^2[\gamma] = \operatorname{Var}[\gamma] / \mathbb E[\gamma]^2$ denotes the square of the coefficient of variation of $\gamma$. 
\end{proposition}

The Pitman relative efficiency $\psi^2$ illuminates how the quality of the proxies $\gamma_i$ affects the power of the weighted test $\phi_\wtd$ relative to the naive test $\phi_\naive$. Inspecting the right-most expression in equation~\eqref{eq:pitman-efficiency}, we can see that the relative efficiency is the ratio between the square of attenuation bias correction factor $\mathbb E[\gamma \mid Z = 1] / \mathbb E[\gamma]$ and the multiplicative increase in variance $1 + \textnormal{c.v.}^2[\gamma]$ brought about by weighting. Therefore, the weighted test gains power over the naive test if and only if its bias reduction outweighs its variance increase. Further inspection of $\psi^2$ gives additional insights. For example, if $\gamma \independent Z$, then we have $\psi^2 = \mathbb E[\gamma]^2 / \mathbb E[\gamma^2] \leq 1$, so the naive test outperforms the weighted test due to the increased variance of the latter; thus, if the weights are uninformative, they should be ignored. Furthermore, provided $\E[\gamma^2 \mid Z=0] < \E[\gamma^2 \mid Z=1]$, it can be seen that $\psi^2$ increases as $\varphi$ decreases. Therefore, the benefit of weighting increases with the frequency of $Z_i = 0$, representing cases where the latent variable $Z_i$ negates the effect $\mu$ in observation $i$. On the other hand, $\psi^2 \leq 1$ as $\varphi \nearrow 1$, so if the latent variables have no impact, then the naive test is best. In sum, which test is more powerful depends on both the rate of the latent variables as quantified by $\varphi$, and the quality of the proxies $\gamma_i$ for the $Z_i$.

\section{Powerful testing in the face of unknown proxy accuracy} \label{eq:testing-unknown-accuracy}

Having established the relationship between the quality of the proxies $\gamma_i$, $\varphi$, and the relative powers of the weighted and naive tests, we arrive at the question of how to ensure powerful testing in practice, when the quality of the proxies and $\varphi$ are unknown. In this section, we propose two approaches to this problem. The first approach (Section~\ref{sec:wtdp}) is based on maximum likelihood estimation (MLE) in a working model, facilitating the adaptive refinement of the proxies based on the data $Y_i$ and thereby improving the power of the weighted test even when the proxies are poor. The second approach (Section~\ref{sec:adapt}) uses auxiliary data to estimate the Pitman relative efficiency $\psi^2$ and adaptively choose whether to use the proxies at all.

\subsection{Improved inference via MLE in a working model}
\label{sec:wtdp}

Intuitively, if the evidence in the observations $Y_i$ runs counter to the proxies $\gamma_i$, we should be able to learn this and adapt the proxies to better fit the data. A natural approach is to employ maximum likelihood estimation (MLE). However, MLE-based inference in the model~\eqref{eq:outcome-model} is cumbersome due to identifiability issues and the fact that the distribution $\gamma_i \mid Z_i$ is left unspecified. A simpler workaround that addresses both issues is to condition on the $\gamma_i$, treating them as fixed. Using the approximation $\mathbb P[Z_i = 1 \mid \gamma_i] \approx \gamma_i$ yields the following \textit{working model}:
\begin{equation}
Z_i \mid \gamma_i \overset{\text{ind}}\sim \textnormal{Ber}(\gamma_i); \quad Y_i \mid Z_i, \gamma_i \overset{\text{ind}}\sim N(\mu Z_i, 1).
\label{eq:working-model}
\end{equation}
Note that this working model coincides with the data-generating model~\eqref{eq:outcome-model} under $H_0: \mu = 0$, which justifies testing within the working model. We can then obtain an estimator $\hat \mu_\wtdp$ of $\mu$ by MLE in this model:
\begin{equation}
\hat \mu_\wtdp \equiv \underset{\mu}{\arg \max}\ \sum_{i = 1}^n \log\left((1 - \gamma_i) \phi(Y_i \mid 0) + \gamma_i \phi(Y_i \mid \mu)\right),
\end{equation}
where $\phi(\cdot \mid \mu)$ denotes the density of a normal distribution with mean $\mu$ and variance 1. In practice, we can fit $\hat \mu_\wtdp$ using the EM algorithm, which involves iterating the following steps until convergence, at which point $\hat\mu = \hat\mu_\wtdp$:
\begin{equation}
\label{eq:EM}
\hat\mu \leftarrow \frac{\sum_{i=1}^n Y_i \hat\gamma_i}{\sum_{i=1}^n \hat\gamma_i} \quad \text{ and } \quad \hat\gamma_i \leftarrow \gamma_i \cdot \frac{ \phi(Y_i \mid \hat\mu)}{(1 - \gamma_i ) \phi(Y_i\mid 0) + \gamma_i \phi(Y_i\mid \hat\mu)}
\end{equation}
Note that the final estimator $\hat\mu_\wtdp$ is also a weighted mean, but unlike $\hat \mu_\wtd$, the weights $\hat \gamma_i$ are adapted to the data. While the working likelihood can be multimodal, this is very unlikely for even moderate $n$ (see Figure \ref{fig:EM-appendix}). We therefore assume henceforth that the output of the EM algorithm is the global maximizer of the working likelihood.

\subsubsection{Testing based on the improved estimate}

In constructing a standard error for $\hat \mu_\wtdp$, we must account for the implicit estimation of the latent variables $Z_i$ to avoid overfitting. To this end, we can use the Fisher-Louis formula \citep{Louis1982a} to get a consistent estimator 
\begin{equation}
\label{eq:fl-se}
\text{s.e.}(\hat \mu_\wtdp) \equiv 1 / \sqrt{\sum_{i=1}^n \hat\gamma_i - \sum_{i=1}^n \hat\gamma_i(1-\hat\gamma_i)(Y_i - \hat\mu_\wtdp)^2}
\end{equation} 
of the standard deviation of $\hat \mu_\wtdp$. We can now define our new weighted test:
\begin{equation}
\label{eq:phi-2}
\phi_\wtdp(\bm Y, \bm \gamma) = \mathbb{I}\left(\frac{\hat \mu_\wtdp}{\text{s.e.}(\hat \mu_\wtdp)} > z_{1 - \alpha}\right).
\end{equation}
As we alluded to previously, the construction of $\phi_\wtdp$ based on the working model~\eqref{eq:working-model} rather than the data-generating model~\eqref{eq:outcome-model} does not impact its Type-I error control. 
\begin{proposition}
\label{eq:asy-calib}
Assume we have i.i.d. samples $(Y_i,\gamma_i)$ generated from model \eqref{eq:outcome-model} and that the regularity conditions in~\eqref{eq:assumptions} hold. Then the test $\phi_\wtdp$ has asymptotic Type-I error control at level $\alpha$.    
\end{proposition}

\subsubsection{Comparison to the weighted test}

Let us compare the original weighted test $\phi_\wtd$ (based on fixed weights) to the proposed weighted test $\phi_\wtdp$ (based on adaptive weights). First suppose $\mu$ is relatively large. If the proxies are good, then intuitively both tests will perform well. But if the proxies are poor, then we would expect the adaptively weighted test to leverage the information in $Y_i$ to improve the weights $\hat\gamma_i$ and thereby outperform the original weighted test. Indeed, our numerical simulations in Section~\ref{sec:simulations} and the appendix support this conclusion. However, if $\mu$ is small, then it becomes unclear whether the variance increase required to accommodate the adaptivity will result in worse power compared to the original weighted test. In fact, as we show in the next proposition, the two tests are asymptotically equivalent in the weak signal strength regime, where $\mu_n = h/\sqrt{n}$.
\begin{proposition}
\label{eq:test1-equals-test2}
Assume we have i.i.d. samples $(Y_i,\gamma_i)$ generated from model \eqref{eq:outcome-model} and that the regularity conditions in~\eqref{eq:assumptions} hold. Then $\mathbb P(\phi_\wtd = \phi_\wtdp) \to 1$ under both the null and local alternatives $\mu_n = h / \sqrt n$ with $h > 0$.  
\end{proposition}

\noindent This result, together with the finite-sample simulation results in Section~\ref{sec:simulations}, leads us to conjecture that the power of $\phi_\wtdp$ dominates that of $\phi_\wtd$ across the full range of signal strengths.

\subsection{Adaptively choosing between weighted and unweighted tests}
\label{sec:adapt}

In some cases, it is best not to use the original weights $\gamma_i$ at all. The ideal way to choose between the weighted test (or its improved variant) and the naive test is to check whether the Pitman relative efficiency $\psi^2$ is greater than 1 (recall Section~\ref{sec:asymptotics}). Of course, $\psi^2$ is unknown in practice, but in this section, we demonstrate how it can be consistently estimated if we have access to a ``positive control outcome variable,'' a variable $Y'$ that shares the same latent variable $Z$ as $Y$, but which is known to have a nonzero mean shift. In other words, $Y' \mid Z \sim N(\mu' Z, 1)$ for $\mu' \neq 0$, and $Y' \independent Y \mid Z$. Positive controls are often available in single-cell CRISPR screens \citep{Papalexi2021}, in the form of a gene known to be impacted by the perturbation under consideration. The use of the positive control also allows us to choose between the tests without ever estimating $\varphi$.

Let us start by recalling from equation~\eqref{eq:pitman-efficiency} the positive square root of the Pitman relative efficiency, 
\begin{equation}
\psi = \frac{\mathbb E[\gamma \mid Z = 1]}{\sqrt{\mathbb E[\gamma^2]}}. 
\end{equation}
The denominator of this quantity is easily estimable from the observed weights $\gamma_i$, but the numerator is more challenging to estimate due to the latent variable $Z$. However, we can note that 
\begin{equation}
\mathbb E[Y'] = \mu' \varphi \quad \text{and} \quad \mathbb E[\gamma Y'] = \mu' \varphi \mathbb E[\gamma \mid Z = 1], \quad \text{so that} \quad \frac{\mathbb E[\gamma Y']}{\mathbb E[Y']} = \E[\gamma\mid Z=1].
\end{equation}
This motivates the definition
\begin{equation}
\label{eq:psi-hat-def}
\hat \psi \equiv \frac{\frac1n\sum_{i=1}^n \gamma_i Y'_i}{\frac1n\sum_{i=1}^n Y'_i \cdot \sqrt{\frac{1}{n} \sum_{i=1}^n \gamma_i^2}}.
\end{equation}
We then immediately have the following:
\begin{proposition}
\label{eq:psi-hat}
Assume we have i.i.d. samples $(Y_i, Y_i',\gamma_i)$ generated from the following augmented variant of our data-generating model~\eqref{eq:outcome-model}:
\begin{equation*}
Z_i \iidsim \textnormal{Ber}(\varphi); \ Y_i \mid Z_i \indsim \mathcal N(\mu Z_i, 1); \  Y'_i \mid Z_i \indsim \mathcal N(\mu' Z_i, 1);\ \gamma_i \mid Z_i \indsim F_{Z_i}; \ Y_i\independent Y'_i \independent \gamma_i \mid Z_i, \ \text{for} \ i = 1, \ldots, n,
\label{eq:augmented-outcome-model}
\end{equation*}
where $\mu' \neq 0$. Then, $\hat\psi \xrightarrow{p} \psi$ with $\hat\psi$ defined as in \eqref{eq:psi-hat-def}.
\end{proposition}

To use $\hat \psi$ within an adaptive test, we must account for the uncertainty of this estimator. While we could obtain an asymptotic limiting distribution for $\hat\psi$, it would involve a very complicated variance estimator. Instead, we use the empirical bootstrap to estimate an asymptotically valid $1-\alpha'$ upper confidence bound for $\psi$, which will be denoted $\hat U_{\alpha'}$. We can then use this upper bound to create tests that adaptively choose between using the weights or not:
\begin{equation}
\label{eq:adaptive-tests}
\phi^a_\wtd \equiv 
\begin{cases}
\phi_\naive & \text{if} \quad \hat U_{\alpha'} < 1, \\
\phi_\wtd & \text{otherwise},
\end{cases}
\quad  \text{and} \quad
\phi^a_\wtdp \equiv 
\begin{cases}
\phi_\naive & \text{if} \quad \hat U_{\alpha'} < 1, \\
\phi_\wtdp & \text{otherwise}.
\end{cases}
\end{equation}
Note that $\hat U_{\alpha'}$ is a function of only $(\bm Y', \bm \gamma)$, whereas the tests $\phi_\naive$, $\phi_\wtd$, and $\phi_\wtdp$ are valid conditionally on $(\bm Y', \bm \gamma)$. This allows the adaptive tests to avoid selection bias and inherit their Type-I error control from the underlying tests. We also note that equation~\eqref{eq:adaptive-tests} uses upper bounds, so we default to the weighted tests and only use the naive tests when we see evidence that the weights are unhelpful. Since $\hat\psi \xrightarrow{p} \psi$, the adaptive tests always use the better test asymptotically. The simulations in Section~\ref{sec:simulations} and the appendix show that these adaptive tests can also improve on the power of their non-adaptive counterparts in finite samples, and that they provide robustness to poor $\gamma$.

\section{Numerical simulations}
\label{sec:simulations}

\begin{figure}[H]
\centering
\includegraphics{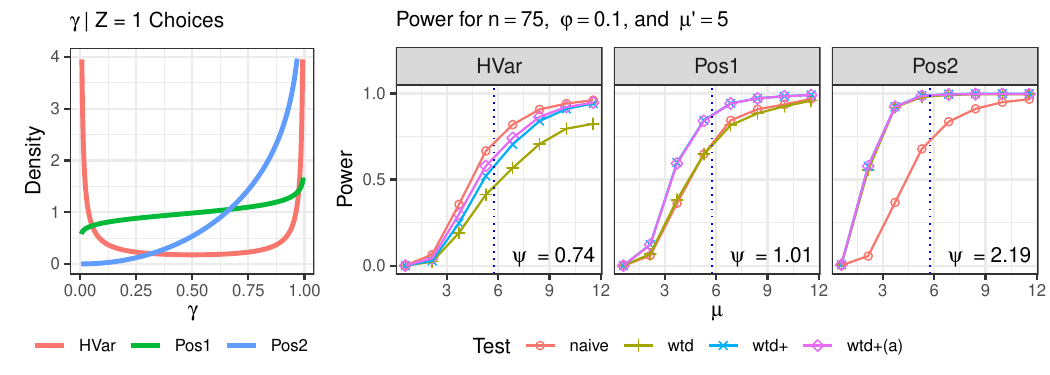}
\caption{Left: The three distributional choices of $\gamma \mid Z = 1$, along with their short-hand names. HVar ("high variance") has $(a,b) = (-0.9, -0.9)$ so $\gamma \independent Z$; Pos1 ("slightly positive") has $(a,b) = (0.1, -0.1)$ so $\gamma$ and $Z$ are slightly positively associated, while Pos2 ("very positive") has $(a,b) = (2, -0.25)$ so $\gamma$ and $Z$ are very positively associated. Right: power as a function of signal strength. $\phi^a_\wtd$ omitted for clarity. Each point reflects the average over 2,000 Monte Carlo repetitions; the maximum standard error is 0.011.}
\label{fig:gammas-and-power}
\end{figure}

In this section, we explore the finite-sample properties of these tests via simulation; code to reproduce all figures is available at \href{https://github.com/Katsevich-Lab/noisy-proxies}{github.com/Katsevich-Lab/noisy-proxies}. We model $\gamma_i$ via
\begin{equation}
\gamma_i \mid Z_i \stackrel{\text{indep.}}\sim \text{Beta}(1 + a Z_i + b(1-Z_i), 1 + a (1-Z_i) + b Z_i), \quad \text{with}\ a, b \in (-1, \infty).
\end{equation}
In this model, $a$ and $b$ are positively and negatively related to the quality of $\gamma_i$, respectively. We choose three sets of parameters $(a,b)$ to represent different levels of association between $\gamma_i$ and $Z_i$; see the left panel of Figure~\ref{fig:gammas-and-power} and Table~\ref{tab:ab-appendix}. All simulations are carried out using a target level of $\alpha = 10^{-4}$ in order to mimic a multiplicity correction as in an actual single-cell CRISPR screen with tens of thousands of perturbation-gene pairs to test for association. 

The right panel of Figure \ref{fig:gammas-and-power} contains our main simulation results. The weighted tests all agree for the largest $\psi$ while differences emerge for smaller $\psi$. $\phi_\wtd$'s power against $\phi_\naive$ is well-described by the asymptotic quantity $\psi$ (recall Proposition~\ref{eq:naive-and-wtd-powers}). $\phi_\wtdp$ matches or improves upon the power of $\phi_\wtd$, a finite-sample effect that complements the asymptotic equivalence in Proposition~\ref{eq:test1-equals-test2}. The adaptive weighted test $\phi_\wtdp^a$ is no worse than $\phi_{\wtdp}$ and offers an improvement when the proxies are poor. Appendix Figures~\ref{fig:calib-appendix}-\ref{fig:asy-expectations} provide additional insights using a wider selection of simulation parameters. In Figure~\ref{fig:calib-appendix}, we verify that all tests control Type-I error at level $\alpha$. The bottom panels of Figure~\ref{fig:power-appendix} demonstrate that the improvement of $\phi_\wtdp^a$ over $\phi_{\wtdp}$ can be more dramatic than in Figure~\ref{fig:gammas-and-power} as the quality of the proxies degrades. Figure~\ref{fig:asy-expectations} demonstrates that $\phi_\wtdp$ can outperform $\phi_\naive$ even when $\psi < 1$, another finite-sample effect that is most prevalent for small $n$ and small $\varphi$. Finally, an increase in $\varphi$ with all else the same leads to a decrease in $\psi$, which tends to favor $\phi_\naive$.

\section{Discussion}

Our analysis answers the two questions posed at the outset. First, the Pitman relative efficiency of the weighted test relative to the naive test provides a natural measure of proxy accuracy, capturing the trade-off between bias reduction and variance inflation. When bias dominates, weighting helps; when added variance dominates, the naive test prevails. Second, we show how to proceed when the proxies’ quality is unknown. An expectation–maximization refinement re-estimates the weights directly from the outcome data, and if a positive-control variable is available, we show how to estimate $\psi^2$ and thereby derive an adaptive means to determine whether to use the proxies. Both of the proposed adaptive tests ($\phi_\wtdp$ and $\phi_\wtdp^a$) outperform the simple weighted test; we recommend applying $\phi_\wtdp$ when a positive control is not available, and $\phi_\wtdp^a$ when it is.

This work opens the door to a broader program of methodological developments aimed at harnessing noisy proxies in latent-variable models and applying these developments to single-cell CRISPR-screen analysis. One strand of future work will generalize our methods to more realistic experimental settings, incorporating two-sample contrasts, count-based and over-dispersed expression distributions, covariate adjustment, and joint modeling of thousands of genes. A second strand will probe and, where needed, strengthen the robustness of the EM-refined and adaptive tests to correlations among the proxy, outcome, and positive-control measurements that arise from shared gene expression patterns. These directions chart a promising path for more powerful inference in single-cell CRISPR screens and other latent variable modeling applications.

\section{Acknowledgments}

E.K. gratefully acknowledges support from NSF DMS-2113072 and NSF DMS-2310654.

% \bibliographystyle{elsarticle-harv}  % for author-year style
% \bibliography{vareff}               % no .bib extension

\begin{small}
\setlength{\bibsep}{0pt}
\bibliographystyle{elsarticle-harv}  % for author-year style
\bibliography{vareff}               % no .bib extension
\end{small}

% \printbibliography  % needed with old Bibtex version

\appendix

\section{Proofs} \label{sec:proofs} % ctrl-f this to get to the start

We will first introduce some notation for convenience. Define $\E_j[\cdot] \equiv \E[\cdot \mid Z = j]$ for $j=0,1$, so e.g. $\E_1[\gamma] = \E[\gamma\mid Z=1]$. Also, define $P_\mu$ to be the probability measure induced by the data-generating model~\eqref{eq:outcome-model} with mean $\mu$.

Next we will give some quantities that will appear throughout these proofs.

\begin{proposition}
\label{eq:moments}
For data $\{(Y_i,\gamma_i) : i=1, \dots, n\}$ generated according to~\eqref{eq:outcome-model} we have the following:

\begin{enumerate}

\item $\E[Y] = \mu \varphi$

\item $\Var[Y] = 1 + \mu^2\varphi(1-\varphi)$

\item $\E[Y\gamma] = \mu \varphi \E_1[\gamma]$

\item $\Var[Y\gamma] = (1-\varphi)\mathbb E_0[\gamma^2] + (\mu^2+1)\varphi\mathbb E_1[\gamma^2] - \mu^2\varphi^2(\mathbb E_1 \gamma)^2$

\item $\Cov[Y\gamma, \gamma] = \mu\varphi\mathbb E_1[\gamma^2] - \mu\varphi \mathbb E_1[\gamma] \mathbb E[\gamma]$

\end{enumerate}

\end{proposition}

In each case the proof follows from applying the law of total expectation with $Z$, and using the conditional independence of $Y$ and $\gamma$ given $Z$ as needed.

\subsection{Proof of Proposition \ref{eq:naive-and-wtd-point-ests}}

\subsubsection{Limiting distribution of $\hat\mu_\naive$ under local alternatives}

\noindent\textbf{Claim:} Under the sequence of local alternatives $\mu_n = h / \sqrt n$ for $h \geq 0$ we have
\begin{equation}
\sqrt n(\hat\mu_\naive - \mu_n \varphi) \xrightarrow{d} \mathcal N(0, 1).
\end{equation}

\noindent\textbf{Proof:} Consider the triangular array $\{X_{nk} : 1\leq k\leq n, n \geq 1\}$ with $X_{nk} = Y_{nk} - \frac h{\sqrt n} \varphi$ where $Y_{n1}, \dots, Y_{nn}$ is an i.i.d. sample from our data-generating process in~\eqref{eq:outcome-model} with mean $\mu_n = h / \sqrt n$. Note that $\E[X_{nk}] = 0$ and $\Var[X_{nk}] = 1 + \frac{h^2}n\varphi(1-\varphi) \to 1 > 0$ as $n\to\infty$. Furthermore, it is easily verified that $\limsup_{n \rightarrow \infty} \mathbb E[|X_{nk}|^{2+\delta}] < \infty$ for any $\delta > 0$, which implies the Lyapunov condition and so the CLT gives us
\begin{equation}
\sqrt n \left(\frac 1n \sum_{k=1}^n X_{nk}\right) = \sqrt n (\hat\mu_\naive - \mu_n \varphi) \xrightarrow{d} \mathcal N(0,1)
\end{equation}
as desired.
\hfill\qed

\subsubsection{Limiting distribution of $\hat\mu_\wtd$ under local alternatives}

\noindent\textbf{Claim:} Assume the regularity conditions in~\eqref{eq:assumptions}. Under the sequence of local alternatives $\mu_n = h / \sqrt n$ for $h \geq 0$ we have
\begin{equation}
\sqrt n \left(\hat\mu_\wtd - \mu_n\varphi\frac{\E_1\gamma}{\E\gamma}\right) \xrightarrow{d} \mathcal N\left(0, \frac{\E[\gamma^2]}{\E [\gamma]^2}\right).
\end{equation}

\noindent\textbf{Proof:} We now consider the vector-valued triangular array $\{\mathbf X_{nk} : 1\leq k\leq n, n \geq 1\}$ with 
\begin{equation}
\mathbf X_{nk} = \left(\begin{array}{c} Y_{nk}\gamma_{nk} - \frac{h}{\sqrt n} \varphi \E_1[\gamma] \\ \gamma_{nk} - \E[\gamma] \end{array}\right)
\end{equation}
where $(Y_{n1}, \gamma_{n1}), \dots, (Y_{nn}, \gamma_{nn})$ is an i.i.d. sample from our data-generating process in~\eqref{eq:outcome-model} with mean $\mu_n = h / \sqrt n$. $\E[\mathbf X_{nk}] = \mathbf 0$ for all $n,k$ and (see Proposition \ref{eq:moments})
\begin{equation}
\Var[\mathbf X_{nk}] = \left(\begin{array}{cc}
(1-\varphi)\mathbb E_0[\gamma^2] + (\mu_n^2+1)\varphi\mathbb E_1[\gamma^2] - \mu_n^2\varphi^2(\mathbb E_1 \gamma)^2 & \cdot \\
\mu_n\varphi\mathbb E_1[\gamma^2] - \mu_n\varphi \mathbb E_1[\gamma] \mathbb E[\gamma] & \Var[\gamma]
\end{array}\right) \to \left(\begin{array}{cc}
\E[\gamma^2] & 0 \\
0 & \Var[\gamma]
\end{array}\right)
\end{equation}
which is positive definite since $\Var[\gamma] > 0$ by assumption. Given that $\gamma_{nk} \in [0,1]$, it is easy to verify that $\limsup_{n \rightarrow \infty} \mathbb E[\|\mathbf X_{nk}]|^{2+\delta}] < \infty$, which again implies the Lyapunov condition. Therefore, the CLT gives us
\begin{equation}
\sqrt n \left(\frac 1n \sum_{k=1}^n \mathbf X_{nk}\right) = 
\sqrt n \left(\frac 1n \sum_{k=1}^n \left[\begin{array}{c}Y_{nk}\gamma_{nk} \\ \gamma_{nk} \end{array}\right] - \left[\begin{array}{c}\mu_n \varphi \E_1[\gamma] \\ \E[\gamma] \end{array}\right]  \right) \xrightarrow{d}\mathcal N\left(\mathbf 0, \left[\begin{array}{cc}
\E[\gamma^2] & 0 \\
0 & \Var[\gamma]
\end{array}\right]\right).
\end{equation}

The centering vector converges to the constant $(0, \E[\gamma])^T$ so we can apply the uniform delta method \citep[Theorem 3.8]{VDV1998} with $g(x,y) = x/y$ to obtain
\begin{equation}
\sqrt n \left( \hat\mu_\wtd - \mu_n \varphi \frac{\E_1[\gamma]}{\E[\gamma]}\right) \xrightarrow{d} \mathcal N\left(0, \frac{\E[\gamma^2]}{\E[\gamma]^2}\right)
\end{equation}
as desired.

\hfill\qed

\subsection{Proof of Proposition \ref{eq:naive-and-wtd-powers}}

\subsubsection{Asymptotic power of $\phi_\naive$ under local alternatives}

\noindent\textbf{Claim:} Under the sequence of local alternatives $\mu_n = h / \sqrt n$ for $h \geq 0$ we have
\begin{equation}
\lim_{n\to\infty} \E[\phi_\naive] = 1 - \Phi(z_{1-\alpha} - h\varphi).
\end{equation}

\noindent\textbf{Proof:} Proposition \ref{eq:naive-and-wtd-point-ests} establishes that $\sqrt n(\hat\mu_\naive - \mu_n\varphi) \xrightarrow{d}\mathcal N(0, 1)$ in this setting. Then
\begin{equation}
\lim_{n\to\infty} \E[\phi_\naive] = \lim_{n\to\infty} \mathbb P\left[\sqrt n \hat\mu_\naive - h\varphi > z_{1-\alpha} - h\varphi\right] = 1 - \Phi(z_{1-\alpha} - h\varphi).
\end{equation}

\hfill\qed

\subsubsection{Asymptotic power of $\phi_\wtd$ under local alternatives} 

\noindent\textbf{Claim:} Assume the regularity conditions in~\eqref{eq:assumptions}. Under the sequence of local alternatives $\mu_n = h / \sqrt n$ for $h \geq 0$ we have
\begin{equation}
\lim_{n\to\infty} \E[\phi_\wtd] = 1 - \Phi\left(z_{1-\alpha} - h\varphi \psi\right)
\end{equation}
where $\psi = \E_1[\gamma] / \sqrt{\E[\gamma^2]}$. 

\noindent\textbf{Proof:} Proposition \ref{eq:naive-and-wtd-point-ests} establishes that $\sqrt n(\hat\mu_\wtd - \mu_n\varphi \E_1[\gamma] / \E[\gamma]) \xrightarrow{d}\mathcal N(0, \E[\gamma^2]/\E[\gamma]^2)$ in this setting. Let $S_n$ be the standard error of $\hat\mu_\wtd$ as given in equation~\eqref{eq:mu-hat-1}, so
\begin{equation}
\sqrt n S_n = \frac{\sqrt{\frac 1n \sum_{i=1}^n \gamma_i^2}}{\frac 1n \sum_{i=1}^n \gamma_i} \xrightarrow{p} \frac{\sqrt{\E[\gamma^2]}}{\E[\gamma]} > 0
\end{equation}
since $\Var[\gamma] > 0$ is assumed. By Slutsky we then have
\begin{equation}
\frac{\sqrt n \left( \hat\mu_\wtd - \mu_n \varphi \frac{\E_1[\gamma]}{\E[\gamma]}\right)}{\sqrt n S_n} = \frac{\hat\mu_\wtd}{\operatorname{s.e.}(\hat\mu_\wtd)} - h \varphi \frac{\E_1[\gamma]}{\sqrt n S_n \E[\gamma]}  \xrightarrow{d} \mathcal N\left(0, 1\right).
\end{equation}
Since $\psi - \frac{\E_1[\gamma]}{\sqrt n S_n \E[\gamma]} = o_{P_n}(1)$ we have
\begin{equation}
% \frac{\hat\mu_\wtd}{\operatorname{s.e.}(\hat\mu_\wtd)} - h \varphi \psi + h \varphi \psi - h \varphi \frac{\E_1[\gamma]}{\sqrt n S_n \E[\gamma]}
\frac{\hat\mu_\wtd}{\operatorname{s.e.}(\hat\mu_\wtd)} - h \varphi \frac{\E_1[\gamma]}{\sqrt n S_n \E[\gamma]} = \frac{\hat\mu_\wtd}{\operatorname{s.e.}(\hat\mu_\wtd)} - h \varphi \psi + o_{P_n}(1)
\end{equation}
so 
\begin{equation}
\lim_{n\to\infty} \E[\phi_\wtd] = \lim_{n\to\infty} \mathbb P\left[\frac{\hat\mu_\wtd}{\operatorname{s.e.}(\hat\mu_\wtd)} - h\varphi\psi > z_{1-\alpha} - h\varphi\psi\right] = 1 - \Phi(z_{1-\alpha} - h\varphi\psi).
\end{equation}

\hfill\qed

\subsection{Proof of Proposition \ref{eq:test1-equals-test2}}

\subsubsection{Main proof}

\noindent\textbf{Claim:} Assume the regularity conditions in~\eqref{eq:assumptions} and that $\hat\mu_\wtdp$ is indeed the MLE of our working model~\eqref{eq:working-model}. Under both the null $H_0 : \mu = 0$ and the sequence of local alternatives $\mu_n = h / \sqrt n$ for $h \geq 0$, we have
\begin{equation}
\mathbb P[\phi_\wtd = \phi_\wtdp] \to 1.
\end{equation}

\noindent\textbf{Proof:} The idea of this proof is to use classical M-estimator theory to connect 
\begin{equation}
T_\wtd = \frac{\hat\mu_\wtd}{\operatorname{s.e.}(\hat\mu_\wtd)} \text{ and } T_\wtdp = \frac{\hat\mu_\wtdp}{\operatorname{s.e.}(\hat\mu_\wtdp)}
\end{equation}
under the null, and to then extend this connection to the sequence of local alternatives $\mu_n = h/\sqrt n$ via contiguity and Le Cam's first lemma. In light of this, all expectations $\E[\cdot]$ in this section are w.r.t. the null distribution. We will present a series of lemmas, which together give the proof of Proposition \ref{eq:test1-equals-test2}. The proofs of these lemmas are deferred to the next section for clarity. 

Define 
\begin{equation}
m_\theta(Y,\gamma) = \log\left((1 - \gamma) \phi(Y\mid 0) + \gamma \phi(Y\mid \theta)\right)
\end{equation}
and let 
\begin{equation}
M_n(\theta) = \frac 1n \sum_{i=1}^n m_\theta(Y_i,\gamma_i).
\end{equation}
Then $M_n$ gives our random objective function and
\begin{equation}
\hat \mu_\wtdp = \underset{\theta}{\arg \max}\ M_n(\theta),
\end{equation}
establishing $\hat\mu_\wtdp$ as an M-estimator. Let 
\begin{equation}
M(\theta) \equiv \E[m_\theta(Y,\gamma)]
\end{equation}
denote the deterministic limiting objective function. The existence of this expectation follows from the proof of Lemma~\ref{eq:uniform-convergence} below. Next, define 
\begin{equation}
\theta^\star \equiv \underset{\theta}{\arg \max}\ M(\theta). 
\end{equation}
We will use $\dot m_\theta(Y,\gamma)$ and $\ddot m_\theta(Y,\gamma)$ for the first two partial derivatives of $\theta \mapsto m_\theta(Y,\gamma)$. Our first step is to compute the first two derivatives of $M$. 
\begin{lemma}
\begin{equation}
M'(\theta) = \mathbb E\left[\dot m_\theta(Y,\gamma)\right] \text{ and } M''(\theta) = \mathbb E\left[\ddot m_\theta(Y,\gamma)\right]
\end{equation}
i.e. the first two derivatives of $M$ can be computed by taking the expectation of the first two partial derivatives of $\theta \mapsto m_\theta(Y,\gamma)$.
\label{eq:exchange}
\end{lemma}

\noindent Next we will establish consistency, in the sense that $\hat\mu_\wtdp \xrightarrow{p} \theta^\star$. The following two lemmas are needed for this:

\begin{lemma}
Under the null $H_0 : \mu = 0$, $\theta^\star = 0$ is the unique maximizer of $M$.
\label{eq:well-separated}
\end{lemma}

\begin{lemma}
$M_n$ converges uniformly in probability to $M$ under the null $H_0 : \mu = 0$, i.e. $\|M_n-M\|_\infty = o_{P_0}(1)$.
\label{eq:uniform-convergence}
\end{lemma}

\noindent Lemmas \ref{eq:well-separated} and \ref{eq:uniform-convergence} are the necessary conditions of Theorem 5.7 in \citet{VDV1998}, which then establishes the consistency of $\hat\mu_\wtdp$ as an M-estimator, so we have $\hat\mu_\wtdp \xrightarrow{p} \theta^\star$ as desired. Continuing, we next need to connect $T_\wtd$ and $T_\wtdp$ under the null. Lemmas \ref{eq:asy-linearity} and \ref{eq:se-consistent} are helper results that allows us to prove Lemma \ref{eq:T1-equals-T2-under-null}, which establishes the desired relationship:

\begin{lemma}
\label{eq:asy-linearity}
Under $H_0: \mu = 0$, $\hat\mu_\wtdp$ admits the following asymptotically linear expansion:
\begin{equation}
\sqrt n \hat\mu_\wtdp = \frac{1}{\mathbb E[\gamma^2] \sqrt n} \sum_{i=1}^n Y_i \gamma_i + o_{P_{0}}(1).
\end{equation}
\end{lemma}

\begin{lemma}
\label{eq:se-consistent}
Under $H_0 : \mu = 0$ we have
\begin{equation}
\operatorname{s.e.}(\sqrt n \hat\mu_\wtdp) \xrightarrow{p} 1 / \sqrt{\E[\gamma^2]}
\end{equation}
\end{lemma}

\begin{lemma}
Under $H_0 : \mu = 0$ we have $T_\wtdp - T_\wtd = o_{P_0}(1)$. 
\label{eq:T1-equals-T2-under-null}
\end{lemma}

\noindent Now that we know that $T_\wtdp - T_\wtd$ vanishes under the null, we need to show that this also happens under our sequence of local alternatives. This will follow from a contiguity argument, which requires that our model is quadratic mean differentiable (QMD).

\begin{lemma}
Our model $\{P_\theta : \theta\in\Theta\}$ is QMD.
\label{eq:qmd}
\end{lemma}

\noindent Lemma \ref{eq:qmd} allows us to apply Theorem 7.2 in \citet{VDV1998}, which gives us
\begin{equation}
\log \prod_{i=1}^n \frac{p_{h/\sqrt n}}{p_0}(\bm Y, \bm \gamma) \xrightarrow{d} \mathcal N\left(-\frac 12 h^2 \E[\gamma^2], h^2 \E[\gamma^2]\right).
\end{equation}

\noindent Then, by Example 6.5 in \citet{VDV1998}, we have $P_{h/\sqrt{n}} \triangleleft \triangleright P_0$, i.e. the laws under the null and the local alternatives are mutually contiguous. Lemma \ref{eq:T1-equals-T2-under-null} established that $T_\wtdp - T_\wtd= o_{P_0}(1)$. Mutual contiguity plus Le Cam's first lemma then gives us $T_\wtdp - T_\wtd= o_{P_{h/\sqrt{n}}}(1)$. These statistics have a Gaussian, and therefore continuous, limiting distribution, and so we have $\mathbb P(\phi_\wtd = \phi_\wtdp)\to 1$ under both the null and local alternatives as desired.

\hfill\qed

\subsubsection{Proofs of supporting lemmas}

We now restate and prove all of the aforementioned lemmas.

\begin{restatelemmawithname}{1}
\begin{equation}
M'(\theta) = \mathbb E\left[\dot m_\theta(Y,\gamma)\right] \text{ and } M''(\theta) = \mathbb E\left[\ddot m_\theta(Y,\gamma)\right]
\end{equation}
i.e. the first two derivatives of $M$ can be computed by taking the expectation of the first two partial derivatives of $\theta \mapsto m_\theta(Y,\gamma)$.
\end{restatelemmawithname}

\begin{proof}
$m_\theta(Y,\gamma)$ is integrable for every $\theta$, and it is twice differentiable w.r.t. $\theta$, so we just need to show that there are integrable dominating functions for each derivative. These will involve polynomials in $|Y|$ which are integrable w.r.t. our Gaussian mixture for any $\mu$.

Computing the first partial derivative, we have
\begin{equation}
\dot m_\theta(Y,\gamma) = \frac{\gamma \phi(Y\mid\theta)(Y-\theta)}{(1-\gamma)\phi(Y\mid 0) + \gamma \phi(Y\mid\theta)}.
\end{equation}
Noting that $(1-\gamma) \phi(Y\mid 0) + \gamma \phi(Y\mid\theta) \geq \gamma \phi(Y\mid\theta)$ we have
\begin{equation}
\left\vert\dot m_\theta(Y,\gamma)\right\vert \leq |Y-\theta| \leq |Y| + K
\end{equation}
thus providing our first dominating function.

Next, define $u(\theta) = \gamma (Y-\theta)\phi(Y\mid\theta)$ and $v(\theta) = (1-\gamma)\phi(Y\mid 0) + \gamma \phi(Y\mid \theta)$ so $\dot m_\theta(Y,\gamma) = \frac{u(\theta)}{v(\theta)}$. Then $u'(\theta) = \gamma \phi(Y\mid\theta)[(Y-\theta)^2-1]$, $v'(\theta) = \gamma (Y-\theta)\phi(Y\mid\theta)$, and
\begin{equation}
\left\vert\ddot m_\theta(Y,\gamma) \right\vert \leq \left\vert \frac{u'(\theta)}{v(\theta)} \right\vert + \left\vert \frac{u(\theta)v'(\theta)}{v(\theta)^2}\right\vert.
\end{equation}
Again use $v(\theta) \geq \gamma \phi(Y\mid\theta)$. This leads to 
\begin{equation}
\left\vert\ddot  m_\theta(Y,\gamma) \right\vert \leq \left\vert [(Y-\theta)^2-1] \right\vert + \left\vert (Y-\theta)^2\right\vert \leq 2(|Y| + K)^2 + 1
\end{equation}
and that gives our second dominating function.

\end{proof}

\begin{restatelemmawithname}{2}
Under the null $H_0 : \mu = 0$, $\theta^\star = 0$ is the unique maximizer of $M$.
\end{restatelemmawithname}

\begin{proof}
For $\theta\neq 0$ we have
\[
\begin{aligned}
M(\theta) - M(0) &= \mathbb{E}\log\left[(1-\gamma)\phi(Y\mid 0) + \gamma \phi(Y\mid \theta)\right] - \mathbb{E} \log \phi(Y\mid 0) \\
&= \mathbb{E}\log\left[\frac{(1-\gamma)\phi(Y\mid 0) + \gamma\phi(Y\mid \theta)}{\phi(Y\mid 0)}\right] \\
&< \log \mathbb{E}\left[\frac{(1-\gamma)\phi(Y\mid 0) + \gamma\phi(Y\mid \theta)}{\phi(Y\mid 0)}\right] \\
&= \log \mathbb{E}_\gamma \mathbb{E}_Y\left[\frac{(1-\gamma)\phi(Y\mid 0) + \gamma\phi(Y\mid \theta)}{\phi(Y\mid 0)}\mid \gamma\right] \\
&= \log \mathbb{E}_\gamma \int_y [(1-\gamma)\phi(Y\mid 0) + \gamma\phi(Y\mid \theta)] \,\text dY \\
&= \log 1 \\
&= 0
\end{aligned}
\]
where the first inequality comes from Jensen's inequality and is strict because we assume $\Var[\gamma] > 0$. This establishes that $\theta^\star = 0$ is the unique maximizer of $M$ under the null.
\end{proof}

\begin{restatelemmawithname}{3}
$M_n$ converges uniformly in probability to $M$ under the null $H_0 : \mu = 0$, i.e. $\|M_n-M\|_\infty = o_{P_0}(1)$.
\end{restatelemmawithname}

\begin{proof}
$M_n$ is an average so this result follows by showing that $\{m_\theta : \theta \in \Theta\}$ is a Glivenko-Cantelli class (see the discussion following Theorem 5.9 in \citet{VDV1998}) $\Theta$ is compact and $\theta \mapsto m_\theta(Y,\gamma)$ is continuous for every $(Y,\gamma)$, so we just need to show that the functions $\theta \mapsto m_\theta$ are dominated by an integrable function uniformly over $\theta$ (again see the discussion following Theorem 5.9 in \citet{VDV1998}). As $\phi(Y\mid\theta) \leq 1 / \sqrt{2\pi}$ for all $Y$ and $\theta$, we will always have $m_\theta(Y,\gamma) < 0$. This means

\[
\begin{aligned}
|m_\theta(Y,\gamma)| &= -\log[(1-\gamma)\phi(Y\mid 0) + \gamma \phi(Y\mid \theta)] \\
&= \log\left[\frac{1}{(1-\gamma)\phi(Y\mid 0) + \gamma \phi(Y\mid \theta)}\right] \\
&\leq -\log[(1-\gamma)\phi(Y\mid 0)] \\
&= -\log(1-\gamma) + \frac{Y^2}{2} + \frac 12 \log(2\pi).
\end{aligned}
\]
We have assumed $-\log(1-\gamma)$ is integrable so this provides a dominating function.

\end{proof}

\begin{restatelemmawithname}{4}
Under the null $H_0: \mu=0$, $\hat\mu_\wtdp$ admits the following asymptotically linear expansion:
\begin{equation}
\sqrt n \hat\mu_\wtdp = \frac{1}{\mathbb E[\gamma^2] \sqrt n} \sum_{i=1}^n Y_i \gamma_i + o_{P_{0}}(1).
\end{equation}
\end{restatelemmawithname}

\begin{proof}
This is a consequence of Theorem 5.23 in \citet{VDV1998}, so we just need to show that the requirements of that theorem are satisfied. First of all, for each $\theta \in (-K,K)$, the map $(Y,\gamma) \mapsto m_\theta(Y,\gamma)$ is measurable, and differentiable at $0$ for a.e. $(Y,\gamma)$ w.r.t. $P_0$. Next, consider two arbitrary points $\theta_1,\theta_2 \in (-K,K)$. We need to show that there exists some $P_0$-square integrable function $\tilde m(Y,\gamma)$ such that
\begin{equation}
|m_{\theta_1}(Y,\gamma) - m_{\theta_2}(Y,\gamma)| \leq \tilde m(Y,\gamma) |\theta_1-\theta_2|.
\end{equation}
In Proposition \ref{eq:exchange} we calculated $\dot{m}_\theta(Y,\gamma) = \frac{\gamma (Y-\theta) \phi(Y\mid\theta)}{(1-\gamma)\phi(Y\mid 0) + \gamma \phi(Y\mid \theta)}$, and we showed that $|\dot{m}_\theta(Y,\gamma)| \leq |Y| + K$. As $Y$ has Gaussian tails, this means $\dot{m}_\theta(Y,\gamma)$ is square integrable w.r.t. $P_0$. Then the following holds by the mean value theorem, for some $\theta^\prime$ between $\theta_1$ and $\theta_2$:
\begin{equation}
\left\vert\frac{m_{\theta_1}(x) - m_{\theta_2}(x)}{\theta_1 - \theta_2} \right\vert = |m_{\theta^\prime}(Y,\gamma)| \leq |Y| + K.
\end{equation}
This shows that we can take $\tilde m(Y,\gamma) = |Y| + K$. Furthermore, Proposition \ref{eq:exchange} shows that $M(\theta)$ is twice differentiable, and we can compute 
\begin{equation}
M^{\prime\prime}(0) = \E\left[\ddot m_{0}(Y, \gamma)\right] = \E[\gamma Y^2 - \gamma - \gamma^2 Y^2].
\end{equation}
We are under the null, so $\gamma \independent Y$ and $\E[Y^2] = 1$, hence $M^{\prime\prime}(0) = -\E[\gamma^2]$. By assumption $\gamma$ is not almost surely $0$, so $M^{\prime\prime}(0) \neq 0$. Finally, we note that Proposition \ref{eq:well-separated} shows that $0$ is the maximum of $M(\theta)$. We have now met the requirements of Theorem 5.23 in \citet{VDV1998}, so we conclude that
\begin{equation}
\sqrt n \hat \mu_n = -\frac{1}{M''(0)}\frac{1}{\sqrt{n}}\sum_{i = 1}^n \dot m_0(Y_i, \gamma_i) + o_{P_0}(1) = \frac{1}{\E[\gamma^2] \sqrt{n}}\sum_{i = 1}^n Y_i \gamma_i + o_{P_0}(1),
\end{equation}
as desired.
\end{proof}

\begin{restatelemmawithname}{5}
Under $H_0 : \mu = 0$ we have
\begin{equation}
\operatorname{s.e.}(\sqrt n \hat\mu_\wtdp) \xrightarrow{p} 1 / \sqrt{\E[\gamma^2]}.
\end{equation}
\end{restatelemmawithname}
\begin{proof}
We prove this directly, rather than via standard MLE machinery, because $\hat\mu_\wtdp$ is the MLE w.r.t. the working model, not the data generating model. Recall
\begin{equation}
\operatorname{s.e.}(\sqrt n \hat\mu_\wtdp) = 1 / \sqrt{\frac 1n \sum_{i=1}^n \left[\hat\gamma_i - \hat\gamma_i(1-\hat\gamma_i)(Y_i-  \hat\mu_\wtdp)^2\right]}.
\end{equation}
The mapping $x \mapsto 1 / \sqrt x$ is continuous for $x>0$, and $\E[\gamma^2] > 0$ by our assumption that $\Var[\gamma] > 0$, so it is sufficient to show
\begin{equation}
\label{eq:obj-initial}
 \frac 1n \sum_{i=1}^n \left[\hat\gamma_i - \hat\gamma_i(1-\hat\gamma_i)(Y_i-  \hat\mu_\wtdp)^2\right] \xrightarrow{p} \E[\gamma^2].
\end{equation}

Lemma \ref{eq:asy-linearity} shows that $\hat\mu_\wtdp = O_p(n^{-1/2})$ under the null, which in turn means $\hat\mu_\wtdp = o_p(1)$. Expanding the square in~\eqref{eq:obj-initial}, we have
\begin{equation}
\label{eq:obj-expanded}
\begin{aligned}
\frac 1n \sum_{i=1}^n \left[\hat\gamma_i - \hat\gamma_i(1-\hat\gamma_i)(Y_i-  \hat\mu_\wtdp)^2\right] &= \frac 1n \sum_{i=1}^n [\hat\gamma_i - \hat\gamma_i(1-\hat\gamma_i)Y_i^2] \\ 
&+\hat\mu_\wtdp \cdot \frac 2n \sum_{i=1}^n \hat\gamma_i(1-\hat\gamma_i) Y_i - \hat\mu_\wtdp^2  \cdot \frac 1n \sum_{i=1}^n \hat\gamma_i(1-\hat\gamma_i).
\end{aligned}
\end{equation}
We have $\hat\gamma_i \in [0,1]$ for each $i$, so $|\hat\gamma_i(1-\hat\gamma_i)| \leq 1$ and therefore
\begin{equation}
\left\vert \frac 1n \sum_{i=1}^n \hat\gamma_i(1-\hat\gamma_i) Y_i \right\vert \leq \frac 1n \sum_{i=1}^n |Y_i|.
\end{equation}
The $Y_i$ are i.i.d. and $\E|Y| < \infty$, so the strong law of large numbers (SLLN) implies $\frac 1n \sum_{i=1}^n |Y_i|$ converges in probability, meaning these partial sums form an $O_p(1)$ sequence. This in turn means $ \frac 1n \sum_{i=1}^n \hat\gamma_i(1-\hat\gamma_i) Y_i = O_p(1)$. We also always have
\begin{equation}
\left\vert \frac 1n \sum_{i=1}^n \hat\gamma_i(1-\hat\gamma_i) \right\vert \leq 1.
\end{equation}
Together these mean
\begin{equation}
\hat\mu_\wtdp \cdot \frac 2n \sum_{i=1}^n \hat\gamma_i(1-\hat\gamma_i) Y_i - \hat\mu_\wtdp^2  \cdot \frac 1n \sum_{i=1}^n \hat\gamma_i(1-\hat\gamma_i) = o_p(1) O_p(1) + o_p(1)O_p(1) = o_p(1),
\end{equation}
so it is therefore sufficient to show
\begin{equation}
\label{eq:obj-simple}
\frac 1n \sum_{i=1}^n [\hat\gamma_i - \hat\gamma_i(1-\hat\gamma_i)Y_i^2] \xrightarrow{p} \E[\gamma^2].
\end{equation}

We now want to express~\eqref{eq:obj-simple} in terms of $\gamma_i$ instead of $\hat\gamma_i$, as that will lead to an average of i.i.d. terms and we can apply the SLLN. To that end, define
\begin{equation}
f_i(\mu) \equiv \frac{\gamma_i \phi(Y_i\mid \mu)}{(1-\gamma_i)\phi(Y_i\mid 0) + \gamma_i \phi(Y_i\mid \mu)}
\end{equation}
so $\hat\gamma_i = f_i(\hat\mu_\wtdp)$ via our EM updates~\eqref{eq:EM}. A first-order Taylor expansion around $0$ gives us
\begin{equation}
\hat\gamma_i = f_i(\hat\mu_\wtdp) = f_i(0) + f_i'(\xi_i)\hat\mu_\wtdp
\end{equation}
for some $\xi_i$ between $\hat\mu_\wtdp$ and $0$. We have $f_i(0) = \gamma_i$, and
\begin{equation}
f_i'(\mu) = \frac{\gamma_i(1-\gamma_i)\phi(Y_i\mid 0)\phi(Y_i\mid \mu)(Y_i - \mu)}{[(1-\gamma_i)\phi(Y_i\mid 0) + \gamma_i \phi(Y_i\mid \mu)]^2}.
\end{equation}
Letting $a = (1-\gamma_i)\phi(Y_i \mid 0)$ and $b = \gamma_i\phi(Y_i\mid \mu)$, we have
\begin{equation}
f_i'(\mu) = \frac{ab}{(a+b)^2} (Y_i - \mu).
\end{equation}
$(a+b)^2 \geq 4ab$, so 
\begin{equation}
|f_i'(\mu)| \leq \frac 14 |Y_i - \mu| \leq |Y_i| + |\mu|.
\end{equation}
As $|\xi_i| \leq |\hat\mu_\wtdp|$, this gives us 
\begin{equation}
|f_i'(\xi_i)| \leq |Y_i| + |\hat\mu_\wtdp|.
\end{equation}
Letting $\hat\gamma_i = \gamma_i + \delta_i$, so $\delta_i = \hat\mu_\wtdp f_i'(\xi_i)$, we have
\begin{equation}
\label{eq:obj-with-deltas}
\begin{aligned}
\frac 1n \sum_{i=1}^n [\hat\gamma_i - \hat\gamma_i(1-\hat\gamma_i)Y_i^2] &= \frac 1n \sum_{i=1}^n [\gamma_i - \gamma_i(1-\gamma_i)Y_i^2] \\
&+ \frac 1n \sum_{i=1}^n \delta_i + \frac 1n \sum_{i=1}^n \delta_i\gamma_i Y_i^2 -  \frac 1n \sum_{i=1}^n \delta_i(1-\gamma_i)Y_i^2 +  \frac 1n \sum_{i=1}^n \delta_i^2 Y_i^2.
\end{aligned}
\end{equation}
The first quantity on the RHS of~\eqref{eq:obj-with-deltas} is our target quantity. If we can show that the 4 summations on the second line of the RHS of~\eqref{eq:obj-with-deltas} are $o_p(1)$, then we are done. This can be done by using $|\delta_i| \leq |\hat\mu_\wtdp|(|Y_i| + |\hat\mu_\wtdp|)$. We will be using the fact that any fixed-degree polynomial of $|Y|$ has finite expectation, so the SLLN applies to i.i.d. averages of such polynomials in $|Y|$.

First we have
\begin{equation}
\begin{aligned}
\left\vert \frac 1n \sum_{i=1}^n \delta_i \right\vert \leq |\hat\mu_\wtdp| \cdot \frac 1n \sum_{i=1}^n( |Y_i| + |\hat\mu_\wtdp|) = o_p(1) [O_p(1) + o_p(1)] = o_p(1).
\end{aligned}
\end{equation}
For the second sum, we have
\begin{equation}
\begin{aligned}
\left\vert \frac 1n \sum_{i=1}^n \delta_i\gamma_i Y_i^2 \right\vert \leq |\hat\mu_\wtdp| \cdot \frac 1n \sum_{i=1}^n( |Y_i| + |\hat\mu_\wtdp|)Y_i^2 = o_p(1) [O_p(1) + o_p(1) O_p(1)] = o_p(1).
\end{aligned}
\end{equation}
The third sum can be treated analogously. Lastly we have
\begin{equation}
\begin{aligned}
\left\vert \frac 1n \sum_{i=1}^n \delta_i^2 Y_i^2 \right\vert \leq \hat\mu_\wtdp^2 \cdot \frac 1n \sum_{i=1}^n( |Y_i| + |\hat\mu_\wtdp|)^2 = o_p(1)[ O_p(1) + o_p(1)O_p(1) + o_p(1)] = o_p(1).
\end{aligned}
\end{equation}
As $o_p(1) + o_p(1) = o_p(1)$, we have established that
\begin{equation}
\plim_{n\to\infty} \frac 1n \sum_{i=1}^n [\hat\gamma_i - \hat\gamma_i(1-\hat\gamma_i)Y_i^2] = \plim_{n\to\infty} \frac 1n \sum_{i=1}^n [\gamma_i - \gamma_i(1-\gamma_i)Y_i^2] .
\end{equation}
These are i.i.d. terms and $\E|\gamma_i - \gamma_i(1-\gamma_i)Y_i^2| < \infty$, so the SLLN applies. Under the null, $\gamma \independent Y$ and $\E[Y^2] = 1$, so 
\begin{equation}
\plim_{n\to\infty} \frac 1n \sum_{i=1}^n [\gamma_i - \gamma_i(1-\gamma_i)Y_i^2] = \E[\gamma - \gamma(1-\gamma)Y^2] = \E[\gamma] - \E[\gamma]\E[Y^2] + \E[\gamma^2]\E[Y^2] = \E[\gamma^2]
\end{equation}
as desired.
\end{proof}

\begin{restatelemmawithname}{6}
Under $H_0 : \mu = 0$ we have $T_\wtdp - T_\wtd = o_{P_0}(1)$. 
\end{restatelemmawithname}

\begin{proof} 
  
By Lemmas~\ref{eq:asy-linearity} and \ref{eq:se-consistent}, we have
\begin{equation}
T_\wtdp = \frac{\hat\mu_\wtdp}{\operatorname{s.e.}(\hat\mu_\wtdp)} = \frac{\sqrt n\hat\mu_\wtdp}{\operatorname{s.e.}(\sqrt n\hat\mu_\wtdp)} = \frac{\frac{1}{\mathbb E[\gamma^2]\sqrt n} \sum_{i=1}^n Y_i\gamma_i + o_{P_0}(1)}{1/\sqrt{\mathbb E[\gamma^2]} + o_{P_0}(1)} = \frac{1}{\sqrt{\mathbb E[\gamma^2]}\sqrt n} \sum_{i=1}^n Y_i\gamma_i + o_{P_0}(1).
\end{equation}
On the other hand, we have
\begin{equation}
T_\wtd = \frac{\hat \mu_\wtd}{\text{s.e.}(\hat \mu_\wtd)} = \frac{\frac{1}{\sqrt{n}}\sum_{i = 1}^n \gamma_i Y_i}{\sqrt{\frac1n\sum_{i = 1}^n \gamma_i^2}} = \frac{1}{\sqrt{\mathbb E[\gamma^2]}\sqrt n} \sum_{i=1}^n Y_i\gamma_i + o_{P_0}(1),
\end{equation}
where the last step follows because $\frac1n\sum_{i = 1}^n \gamma_i^2 \xrightarrow{p} \mathbb E[\gamma^2]$ by the SLLN and $\frac{1}{\sqrt n} \sum_{i=1}^n Y_i\gamma_i$ converges to a normal limit by the CLT and therefore $\frac{1}{\sqrt n} \sum_{i=1}^n Y_i\gamma_i = O_{P_0}(1)$. Comparing the last two displays yields the desired conclusion.
\end{proof}

\begin{restatelemmawithname}{7}
Our model $\{P_\theta : \theta\in\Theta\}$ is QMD.
\end{restatelemmawithname}

\begin{proof}

This is a consequence of Lemma 7.6 in \citet{VDV1998}, so the result follows from showing the conditions of this lemma are satisfied. Let $p_\theta$ denote the Lebesgue density of $P_\theta$. The first requirement is that, for all $\theta \in (-K,K)$, the map $\theta \mapsto \sqrt{p_\theta(Y,\gamma)}$ is continuously differentiable for all $(Y,\gamma)$. This is indeed true for our particular data generating process.

The second and final requirement is that $I_\theta := \int (\dot p_\theta / p_\theta)^2 p_\theta \,\text d\gamma \text d y$ is well-defined and continuous in $\theta$. Suppose that there is an integrable function $g(\gamma, Y)$ that dominates $(\dot p_\theta / p_\theta)^2 p_\theta$. Then $I_\theta$ is well-defined and finite, and if it is also the case that $(\dot p_\theta / p_\theta)^2 p_\theta$ is continuous in $\theta$, then the dominated convergence theorem establishes the continuity of $I_\theta$. 

We have
\begin{equation}
\begin{aligned}
I_\theta &= \int \left(\frac{\varphi g_1(\gamma)\phi(y-\theta)}{p_\theta(y,\gamma)}\right)^2 (y-\theta)^2 p_\theta(y,\gamma)\,\text dy\,\text d\gamma \\
&= \int \mathbb P(Z=1\mid y,\gamma,\theta,\varphi)^2 (y-\theta)^2 p_\theta(y,\gamma)\,\text dy\,\text d\gamma.
\end{aligned}
\end{equation}
This integrand is continuous in $\theta$, so we just need to produce a dominating function and we are done. To that end, let $g_0(\gamma)$ and $g_1(\gamma)$ denote the Lebesgue densities of $\gamma \mid Z=0$ and $\gamma \mid Z=1$, respectively. Then
\begin{equation}
p_\theta(y,\gamma) = (1-\varphi)g_0(\gamma) \phi(y) + \varphi g_1(\gamma)\phi(y-\gamma)
\end{equation}
and $\mathbb P(Z=1\mid Y,\gamma) \leq 1$, so we immediately get an upper bound of
\begin{equation}
(y-\theta)^2 [g_0(\gamma)\phi(y) + g_1(\gamma)\phi(y-\theta)].
\end{equation}
Next, because $y-\theta \leq |y| + |\theta| \leq |y| + K$, we have $(y-\theta)^2 \leq (|y| + K)^2$. Furthermore, let
\begin{equation}
h(y) = \begin{cases}
\phi(y +K) & y < -K \\
1/\sqrt{2\pi} & -K \leq y \leq K \\
\phi(y - K) & y > K
\end{cases}
\end{equation}
so $h$ is constant on $\Theta$ and decays as a Gaussian on either side. Then $h(y) \geq \phi(y-\theta)$ for any $\theta \in \Theta$ (pf: consider the three cases corresponding to $y < -K$, $y\in \Theta$, and $y > K$). This means we have
\begin{equation}
(|y| + K)^2 [g_0(\gamma)\phi(y) + g_1(\gamma) h(y)]
\end{equation}
as our bounding function, and this is integrable and independent of $\theta$. Thus the requirements of Lemma 7.6 in \citet{VDV1998} are satisfied, and $\{P_\theta : \theta \in \Theta\}$ is QMD.
\end{proof}

\section{Additional simulations} \label{sec:sims-appendix}

\subsection{Typical Log Likelihood}

\begin{figure}[htpb]
\centering
\includegraphics{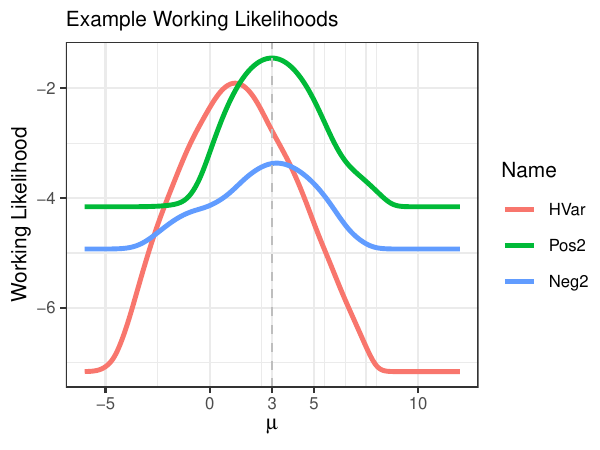} %[width=0.8\textwidth]
\caption{Typical realized log likelihoods, justifying our assumption that the EM algorithm will return the global optimum and $\hat\mu_\wtdp$ will actually be the MLE in practice. These simulations use $n = 10$, $\varphi = 0.3$, and $\mu = 3$. See Table \ref{tab:ab-appendix} for the precise $\gamma$ distribution parameters.}
\label{fig:EM-appendix}
\end{figure}

In Section~\ref{sec:wtdp} we observed that the log likelihood given by $M_n$ has the possibility of being multimodal. Here in Figure \ref{fig:EM-appendix} we give simulated examples this likelihood. Even for this modest sample size of $n=10$ each likelihood is unimodal with a clear mode near $\mu$, thus supporting our assumption that $\hat\mu_\wtdp$ is the actual MLE.

\subsection{Additional calibration and power simulations}

Next we present additional simulations exploring the Type-I error control and power of these tests. Table \ref{tab:ab-appendix} gives the values of $a$ and $b$ used here.

\begin{table}[H]
\centering
\caption{Choices for $(a,b)$ in the distribution of the $\gamma_i$ along with the abbreviations used in the supplementary simulations.}
\label{tab:ab-appendix}
\begin{tabular}{|l|c|c|r|}
\hline
\textbf{Description} & \textbf{Abbreviation} & \textbf{$a$} & \textbf{$b$} \\
\hline
High variance & HVar & -0.9 & -0.9 \\
Uniform & Unif & 0 & 0 \\
% Concentrated & Conc & 4 &  4 \\
Slightly positive & Pos1 & 0.1 & -0.1 \\
Positive & Pos2 & 2 & -0.25 \\
% Very positive & Pos3 & 3.8 & -0.7 \\
Slightly negative & Neg1 &  -0.1 & 0.1 \\
Negative & Neg2 &  -0.25 & 2 \\
% Very negative & Neg3 &  -0.9 & 4 \\
\hline
\end{tabular}
\end{table}

\begin{figure}[htpb]
\centering
\includegraphics{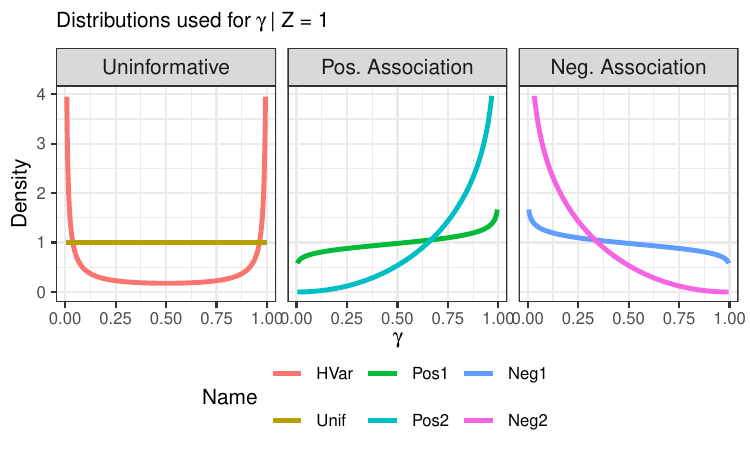} %[width=0.8\textwidth]
\caption{The 6 different distributions of $\gamma_i$ that are used for our simulations in the additional simulations. See Table \ref{tab:ab-appendix} for the specific choices of $a$ and $b$.}
\label{fig:gammas-appendix}
\end{figure}

Figure \ref{fig:gammas-appendix} shows the resulting distributions.

\subsubsection{Calibration}

\begin{figure}[htpb]
\centering
\includegraphics[width=0.95\textwidth]{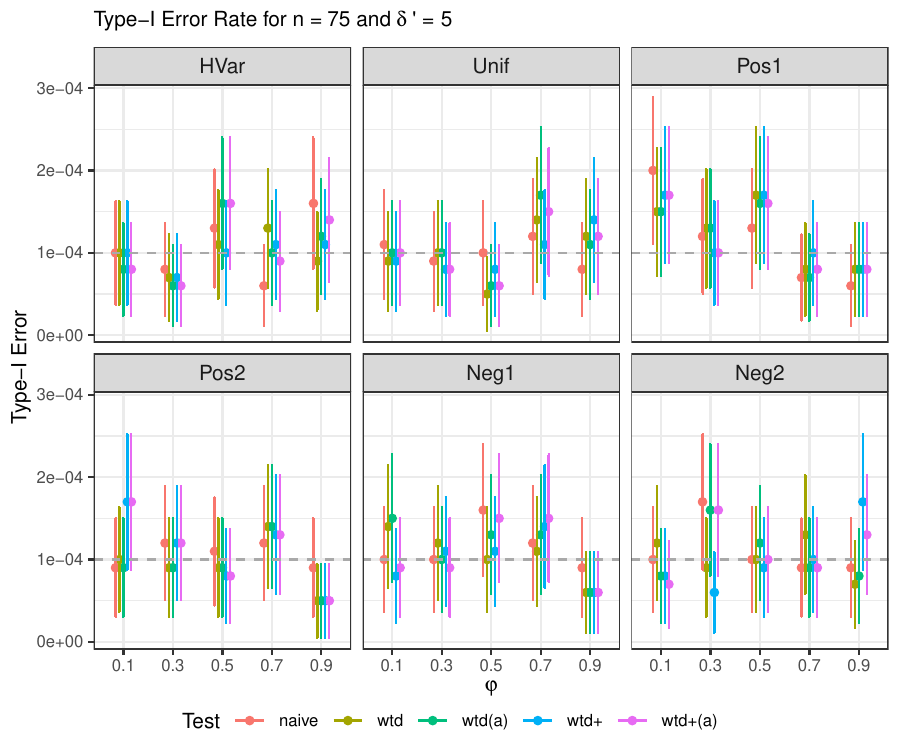}
\caption{This figure gives the simulated Type-I error rates of these tests. The dashed horizontal line shows the nominal Type-I error rate of $10^{-4}$. The positive control mean is $\mu' = \frac{\delta'}{\varphi \sqrt{n}}$. We see no evidence of a lack of Type-I error control.}
\label{fig:calib-appendix}
\end{figure}

Figure \ref{fig:calib-appendix} shows that these tests keep Type-I error control for this larger collection of distributional choices for $\gamma$. 

\subsubsection{Power}

\begin{figure}[htpb]
\centering
\includegraphics[width=0.95\textwidth]{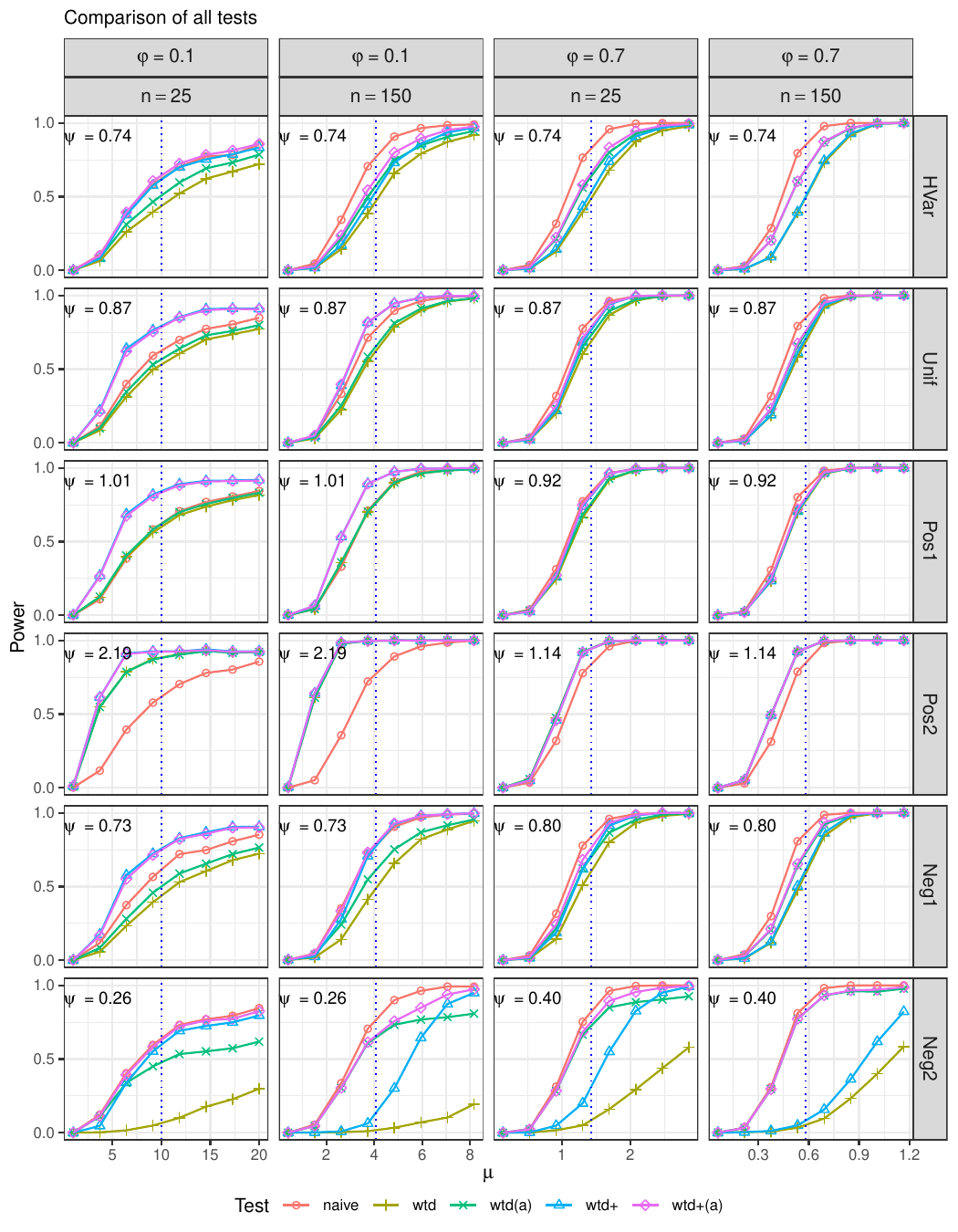}
\caption{The power of all five tests are compared for $n = 75$. The dotted vertical line indicates $\mu'$. For this figure $\mu'$ is chosen so that it corresponds to the middle of the sequence of $\mu$ that produce non-trivial powers. The maximum Monte Carlo standard error for any point is 0.011. The conclusions are the same as in Figure \ref{fig:gammas-and-power}.}.
\label{fig:power-appendix}
\end{figure}

Figure \ref{fig:power-appendix} explores the power of these tests in this expanded setting.

\begin{figure}[htpb]
\centering
\includegraphics{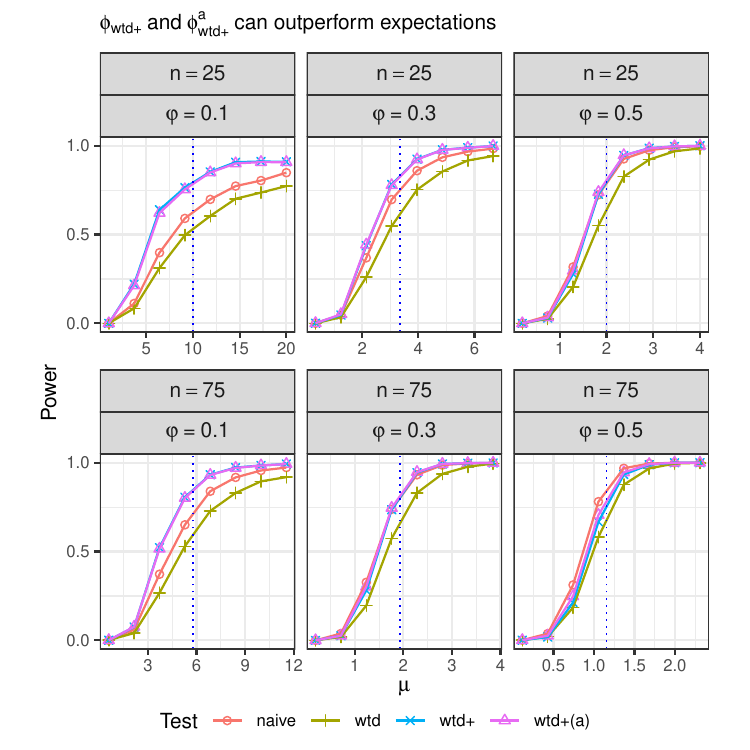}
\caption{For small $n$ and $\varphi$, the tests $\phi_\wtdp$ and $\phi_\wtdp^a$ can be more powerful than $\phi_\naive$ even though this is not the case asymptotically. The advantage disappears as either $n$ or $\varphi$ increase. In this figure we have $\gamma \sim \text{Unif}(0,1)$ so $\psi \approx 0.87$. The maximum standard error for any point is 0.011.}
\label{fig:asy-expectations}
\end{figure}

Figure \ref{fig:asy-expectations} explores how $\phi_\wtdp$ can outperform what we'd expect from our asymptotics, whereas $\phi_\wtd$ is well-described by the same asymptotics.

\end{document}